\documentclass[aps,prb,twocolumn,showpacs,superscriptaddress]{revtex4-1}
\usepackage{epsfig}
\usepackage[T1]{fontenc}
\usepackage[latin9]{inputenc}
\usepackage{color}
\usepackage{amsmath, amsthm, amssymb}
\usepackage[latin9]{inputenc}
\usepackage{hyperref}
\usepackage{graphicx}
\usepackage{subfigure}

\include{mydefs}

\begin{document}


\title{Evolution of the Superconductivity Dome in the two dimensional Hubbard Model}


\author{K.-S. Chen}
\email{kuangshingchen@gmail.com}
\affiliation{Department of Physics and Astronomy, Louisiana State University, Baton Rouge, LA 70803, USA}
\author{Z. Y. Meng}
\email{ziyangmeng@gmail.com}
\affiliation{Department of Physics and Astronomy, Louisiana State University, Baton Rouge, LA 70803, USA}
\affiliation{Center for Computation and Technology, Louisiana State University, Baton Rouge, LA 70808, USA}
\author{S.-X. Yang}
\affiliation{Department of Physics and Astronomy, Louisiana State University, Baton Rouge, LA 70803, USA}
\affiliation{Center for Computation and Technology, Louisiana State University, Baton Rouge, LA 70808, USA}
\author{T. Pruschke}
\affiliation{Department of Physics, University of G\"ottingen, D-37077 G\"ottingen, Germany}
\author{J. Moreno}
\affiliation{Department of Physics and Astronomy, Louisiana State University, Baton Rouge, LA 70803, USA}
\affiliation{Center for Computation and Technology, Louisiana State University, Baton Rouge, LA 70808, USA}
\author{M. Jarrell}
\affiliation{Department of Physics and Astronomy, Louisiana State University, Baton Rouge, LA 70803, USA}
\affiliation{Center for Computation and Technology, Louisiana State University, Baton Rouge, LA 70808, USA}

\date{\today}

\begin{abstract}
In a recent publication {[}Chen \textit{et al}., Phys. Rev. B \textbf{86}, 165136 (2012){]}, 
we identified a line of Lifshitz transition points separating the Fermi liquid and pseudogap regions 
in the hole-doped two dimensional Hubbard model. Here we extend the study to further determine 
the superconducting transition temperature in the phase diagram. By means of large-scale dynamical 
cluster quantum Monte Carlo simulations, we are able to identify the evolution of the $d$-wave 
superconducting dome in the hole-dope side of the phase diagram, with next-nearest-neighbor hopping ($t'$),
 chemical potential and temperature as control parameters. To obtain the superconducting transition 
temperature $T_c$, we employ two-particle measurements of the pairing susceptibilities. As $t'$ goes 
from positive to negative values, we find the $d$-wave projected irreducible pairing vertex function 
is enhanced, and the curvature of its doping dependence changes from convex to concave, which fixes 
the position of the maximum superconducting temperature at the same filling ($n\approx0.85$) and 
constraints the dome from precisely following the Lifshitz line.   We furthermore 
decompose the irreducible vertex function into fully irreducible, charge 
and spin components via the parquet equations, and consistently find that the spin component dominates 
the pairing vertex function in the doping range where the dome is located. Our investigations 
deepen the understanding of the phase diagram of the two dimensional Hubbard model, and more importantly 
pose new questions to the field. For example, we found as $t'$ goes from positive to negative values, the 
curvature of the pairing strength as a function of doping changes from convex to concave, and the nature of the
dominant fluctuations changes from charge degree of freedom to spin degree of freedom.
The study of these issues will lead to further understanding 
of the phase diagram of the two dimensional Hubbard model and also the physics of the hole-doped cuprate 
high temperature superconductors.
\end{abstract}

\pacs{74.40.Kb, 71.10.Fd, 74.72.-h, 71.10.Hf}

\maketitle

\section{Introduction}
The phase diagram of the hole-doped two-dimensional Hubbard model in the strongly correlated regime 
contains many interesting features resembling those observed in the high-$Tc$ cuprate superconductors~\cite{Norman2013,Anderson2004}.
At a qualitative level, it is accepted that the generic ingredients of the cuprate physics are 
captured by the one band Hubbard model on the square lattice~\cite{Phillips2010,Alloul2013}. For example, 
the Fermi surface reconstruction is observed in the study of the Hubbard model on the square lattice as 
the doping concentration is varied~\cite{Chen11,Chen12}.  Starting from the overdoped Fermi liquid, with 
an electron-like Fermi surface, quadratic resistivity in temperature and well-defined quasiparticles, 
the system exhibits around optimal doping for the temperature regime studied a vanishing quasiparticle 
weight on the Fermi surface and a linear resistivity, reminiscent of a marginal Fermi liquid~\cite{Varma96,Varma99}.
At smaller doping concentrations, the system enters a pseudogap region 
where the density of states is further suppressed close to the Fermi level, and the topology of the Fermi surface 
changes to hole-like. Eventually at half-filling, a Mott gap opens and antiferromagnetic 
order sets in.

Among the many interesting questions concerning the phase diagram of the hole-doped Hubbard model 
on a square lattice, the presence of the $d$-wave superconducting dome, as observed in the 
cuprate superconductors, is the central one. It is commonly accepted that the physics 
of cuprates is due to moderate to strong electronic correlations. Therefore, any study of 
low-temperature properties faces the difficulty that one cannot employ conventional perturbative
 methods to obtain reliable information. In this paper, we address the existence of a $d$-wave 
superconductivity dome in the Hubbard model directly using large-scale dynamical cluster 
quantum Monte Carlo simulations~\cite{Hettler98,Hettler00,Jarrell01,Maier05}. Based on the 
understanding obtained in this and previous numerical works
~\cite{Marcridin06,Vidhyadhiraja09,Mikelsons2009,Liebsch09,Khatami10,Lin10,Gull10,Sakai10,YangSX11,Galanakis11,Chen11,Chen12,Sordi2012,Sakai12,Gull2013}, 
we furthermore map the evolution of the $d$-wave superconductivity dome in the parameter space of the phase diagram.

In order to put this work into the proper perspective, it is worthwhile to recapitulate our knowledge 
about the phase diagram of the two-dimensional Hubbard model. Using large-scale dynamical cluster 
quantum Monte Carlo simulations, the phase diagram of the two-dimensional Hubbard model near 
quantum critical filling (optimal filling or doping) was mapped out in a series of numerical works
~\cite{Marcridin06,Vidhyadhiraja09,Mikelsons2009,Khatami10,YangSX11,Galanakis11,Chen11,Chen12}. 
The control parameters are the next-nearest-neighbor hopping ($t'/t$), chemical potential 
(or doping concentration) and temperature. At positive $t'$, there is a first-order phase separation 
transition occurring at finite temperature. The two phases being separated are an incompressible 
Mott liquid and a compressible Mott gas; these two phases are adiabatically connected to the pseudogap 
and the Fermi liquid states at $t'=0$. The first order line of coexistence terminates 
at a second order point where the charge susceptibility diverges~\cite{Marcridin06,Mikelsons2009,Khatami10,Galanakis11}. 
As $t'\rightarrow0$, this critical point extrapolates continuously to zero temperature and thus 
becomes the quantum critical point (QCP) underneath the superconducting dome~\cite{YangSX11}. 
Above the QCP, a V-shaped quantum critical region separates the pseudogap and the Fermi liquid phases
~\cite{Vidhyadhiraja09,Chen11}. 

When the next-nearest-neighbor hopping becomes negative, $t'<0$, which is relevant for the hole-doped cuprates, 
the physics becomes more interesting~\cite{Chen11,Chen12}. We find at $t'/t \le 0$, as the doping concentration 
varies from the overdoped to the underdoped regime, the Fermi surface changes its topology from electron-like 
with complete Fermi surface to hole-like with a pseudogap in the anti-nodal direction. Such a topological transition 
in the Fermi surface is a Lifshitz transition~\cite{Lifshitz60,Yamaji06,Imada2010,Imada2011,Chen12}. 
It is furthermore concomitant with a van Hove singularity in the density of states crossing the Fermi 
level at a doping which occurs very close to (if not at) the quantum critical point. 
Interestingly, we find the quantum critical phenomena prevail along the line of Lifshitz transition for 
negative $t'$.  The temperature dependence of the correlation effects close to the van Hove singularities, 
and its influence on quantities like the quasiparticle fraction and the pairing polarization 
are very different from those found in the traditional van Hove scenarios~\cite{Hirsch86b,Markiewicz97}. 

A schematic phase diagram based on these studies is provided in Fig. $8$ of Ref.~\onlinecite{Chen12}. 
However, the position of the superconducting dome and its evolution in the negative $t'$ side 
of the phase diagram have not been addressed. From our previous work~\cite{YangSX11,Chen12} we understand 
that the formation of the $d$-wave superconducting dome is due to two competing factors: the effective 
pairing strength $V_d$ and the $d$-wave pairing bubble $\chi^{0}_{d}$. When $t'=0$, $V_d$ decreases 
monotonically as a function of doping while the $\chi^{0}_{d}$ shows algebraic divergence 
($\displaystyle \frac{1}{\sqrt{T}}$) near the quantumcritical doping (Lifshitz doping) and 
a much slower dependence in the pseudogap and Fermi liquid regions. The combined effect of the monotonic 
decay of $V_d$ and the algebraic divergence of $\chi^{0}_{d}$ results in the maximum of the dome occurring 
near the critical doping. 

In this paper we extend the study of the $d$-wave pairing and the superconducting dome to the negative $t'$ 
side of the phase diagram. To this end, we include extensive two-particle measurements into the dynamical 
cluster quantum Monte Carlo simulations. This allows to measure the $d$-wave pairing susceptibility directly, 
and obtains the superconducting transition temperature $T_c$ from its divergence. We find that the superconducting dome 
is located around the quantum critical doping, and furthermore follows its evolution into negative $t'$ values. 
The evolution of the dome is subtle, in the sense that its peak stays close to the same filling ($n \approx 0.85$), 
whereas the whole dome moves outside towards higher doping, following the direction of the Lifshitz line. The 
$d$-wave projected irreducible pairing vertex function (the effective pairing strength $V_d$) is enhanced as $t'$ 
changes from positive to negative values, in the under- and optimal-doped regions.  The doping dependence of the 
effective pairing strength, changes its curvature from convex to concave as $t'$ goes to negative values, which 
fixes the position of the maximum superconducting temperature (the peak of the dome) at the same filling ($n\approx0.85$) and 
constraints the dome from precisely following the Lifshitz line. Therefore, the asymmetry of the superconducting dome 
becomes more  pronounced when $t'$ goes to negative values due to the non-trivial doping dependence of $V_d$. 
The vertex decomposition via parquet equations furthermore reveals that in the negative $t'$ side of the phase diagram, 
the effective pairing strength is dominated 
by magnetic fluctuations associated with the antiferromagnetic order at momentum transfer 
$\mathbf{Q}=(\pi,\pi)$.

The paper is organized as follows. Section II outlines the model and the methods used in this study: 
the dynamical cluster approximation (DCA) with weak-coupling continuous-time quantum Monte Carlo (CTQMC)
 as its cluster solver. We explain in detail how the two-particle vertex function and its decomposition 
has been included into our DCA/CTQMC simulations. Section III and IV contain our numerical results 
and discussion, beginning with $d$-wave pairing susceptibility and the evolution of the superconducting dome
 as a function of $t'/t$, and followed by a detailed account of the irreducible pairing vertex function 
and its decomposition via parquet equations. We then provide results on the cluster spin susceptibility 
at a fixed filling and various $t'/t$. We use a schematic quantum critical phase diagram of the model to 
summarize our results. We end with conclusions and an overview of open questions in Section V.

\section{Formalism}
\label{sec:formalism}
The model we study in this paper is the Hubbard model on a square lattice  
\begin{equation}
H=\sum_{\mathbf{k}\sigma}\left(\epsilon_{\mathbf{k}}^{0}-\mu\right)c_{{\mathbf{k}}\sigma}^{\dagger}
c_{{\mathbf{k}}\sigma}^{\phantom{\dagger}}+U\sum_{i}n_{{i}\uparrow}n_{{i}\downarrow},
\label{eq:hubbard}
\end{equation}
where $c_{{\mathbf{k}}\sigma}^{\dagger}(c_{{\mathbf{k}}\sigma})$ is the creation (annihilation) operator for 
electrons with wavevector ${\mathbf{k}}$ and spin $\sigma$, $\mu$ is the chemical potential, $n_{i\sigma} =c_{i\sigma}^{\dagger}c_{i\sigma}$ is the number operator, $U$ is the on-site Coulomb repulsion, and the bare dispersion is given by
\begin{equation}
\epsilon_{\mathbf{k}}^{0}=-2t\left(\cos k_{x}+\cos k_{y}\right)-4t'\cos k_{x}\cos k_{y},
\label{StdDisp}
\end{equation}
with $t$ and $t'$ being the hopping amplitudes between nearest and  next-nearest-neighbor sites 
respectively.

We use the DCA~\cite{Hettler98, Hettler00} with interaction-expansion CTQMC~\cite{Rubtsov05} 
as a cluster solver. The DCA is a mean-field theory on clusters that maps the lattice 
of the original system onto a periodic cluster of size $N_c=L_c^D$ ($D$ is the dimensionality) 
embedded in a self-consistently determined host. The spatial short-ranged correlations 
(up to $L_c$) are treated explicitly while the long-ranged correlations are taken into account 
in a mean-field approximation. In this paper we study cluster sizes of $N_c=12$ and $16$, 
and compared results with other cluster sizes in previous studies~\cite{Maier06,Vidhyadhiraja09,Mikelsons2009,Khatami10,Galanakis11,YangSX11,Chen11,Chen12}. 
The energy unit is $4t=1$, and the interaction strength is set as $U=6t$. 
The interaction-expansion CTQMC solver treats the temporal correlations, especially important 
for quantum criticality, for all cluster sizes. In our simulation, we can achieve temperatures
 as low as $T=0.04t$ ($\beta t= 22.5$) before a serious minus-sign problem renders the data untrustable.

\begin{figure}[h!]
\centerline{
\includegraphics[width=3.3in]{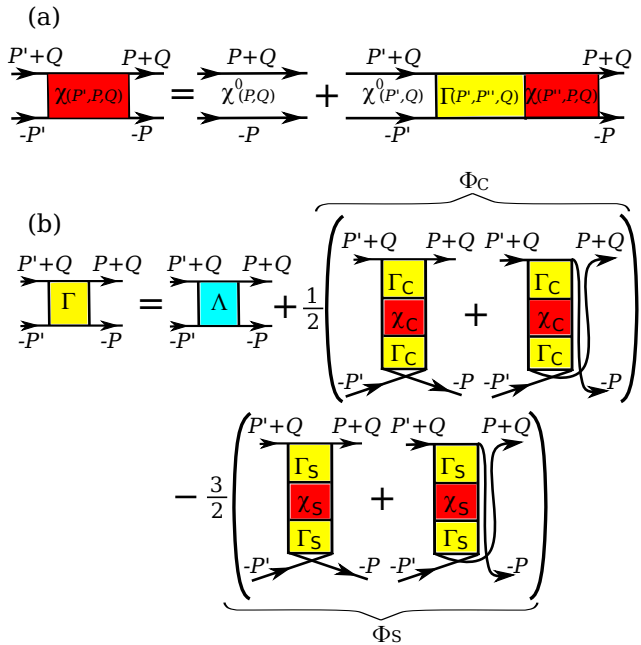}}
\caption{(color online) (a) Bethe Salpeter equation for the particle-particle pairing channel. 
$\chi(P,P',Q)$ is the two particle Green function and $\Gamma(P',P",Q)$ is the irreducible vertex function, 
both are defined on the DCA/CTQMC cluster. $\chi^{0}(P,Q)$ is the bare bubble coarse-grained 
from lattice to cluster. (b) Decomposition of the irreducible particle-particle vertex $\Gamma$ 
into a fully irreducible vertex $\Lambda$ and cross channel contributions from charge $\Phi_C$ 
and spin $\Phi_S$ components. Both $\Phi_C$ and $\Phi_S$ come from the particle-hole channel 
with $\chi_{S/C}$ and $\Gamma_{C/S}$ as the corresponding two-particle Green functions 
and irreducible vertex functions extracted from their own Bethe Salpeter equations, similarly as that in (a).}

\label{fig:BetheSalpeterEq}
\end{figure}

The self-consistent loop of the DCA is built at the single-particle level, hence we can obtain 
the lattice Green function $G(\mathbf{k},i\omega_{n})$ and self-energy $\Sigma(\mathbf{k},i\omega_{n})$,
and perform their analytical continuation using the Maximum Entropy Method (MEM)~\cite{Jarrell96,Wang09,Chen11,Chen12} 
to obtain the real-frequency
 single-particle quantities such as $G(\mathbf{k},\omega)$, $\Sigma(\mathbf{k},\omega)$, local density 
of states $N(\omega)$ and spectral function $A(\mathbf{k},\omega)$. However, to obtain various 
susceptibilities and vertex functions, we need to go to the two-particle level in our DCA/CTQMC simulations.
To this end, we measure the two-particle Green function, $\chi(P,P',Q)$, where the four-momentum notation 
$P\equiv(\mathbf{K},i\omega_n)$ and $Q\equiv(\mathbf{Q},i\nu_m)$ are introduced, with $\mathbf{K}$, $\mathbf{Q}$ 
are momentum points on the cluster, and $\omega_n=(2n+1)\pi T$ and $\nu_m=2m\pi T$ are fermionic 
and bosonic Matsubara frequencies, respectively. From the two-particle Green function, we extract 
the irreducible vertex function, $\Gamma(P,P',Q)$,  via the Bethe-Salpeter equation. Fig.~\ref{fig:BetheSalpeterEq}
 (a) demonstrates an example of the Bethe-Salpeter equation in the particle-particle pairing channel. 
The bare bubble $\chi^{0}(P,Q)=G(-P)G(P+Q)$ entering the Bethe-Salpeter equation is obtained 
from the fully dressed single-particle Green function, coarse-grained from lattice to cluster. 
Once we have the irreducible vertex, $\Gamma$, the lattice susceptibility can be calculated 
from $\chi(T)=\displaystyle \frac{\chi^0}{1-\Gamma\chi^0}$, where the summation is over all the frequency 
and momentum indices. Since the square lattice Hubbard model exhibits $d$-wave superconductivity, 
the lattice $d$-wave susceptibility, $\chi_{d}(T)$, is obtained by projecting the pairing susceptibility 
with the $d$-wave form factor 
$g_d(\mathbf{k})=\cos(k_x)-\cos(k_y)$ via  $\chi_{d}(T)=\sum_{\mathbf{k},\mathbf{k'}}g_d(\mathbf{k})\chi_{\mathbf{k},\mathbf{k'}}(T)g_d(\mathbf{k'})/\sum_{\mathbf{k}}g^{2}_{d}(\mathbf{k})$. 
The superconducting transition temperature can be determined by extrapolating $1/\chi_d(T)$ to zero. 

To further explore the pairing mechanism, we decompose the irreducible pairing vertex $\Gamma$, 
by means of the parquet equation~\cite{n_bulut_93,Maier06,SXYang2009,YangSX11,Chen2013}, into three independent contributions, 
$\Gamma=\Lambda+\Phi_C+\Phi_S$, as shown in Fig.~\ref{fig:BetheSalpeterEq} (b). These three contributions 
are the fully irreducible vertex $\Lambda$, the charge ($S=0$) component $\Phi_C$, and the spin ($S=1$) 
component $\Phi_S$. We further project both sides of the equation with the $d$-wave form factor. 
For example, the effective $d$-wave pairing strength, is given by 
$V_d=\sum_{\mathbf{k},\mathbf{k'}}g_{d}(\mathbf{k})\Gamma_{\mathbf{k},\mathbf{k'}}g_{d}(\mathbf{k'})/\sum_{\mathbf{k}}g^{2}_{d}(\mathbf{k})$.
 Such vertex decomposition approach tells which component contributes most to the irreducible pairing vertex. 
One important point to note is that as we have controlled information about the two particle vertex function
 in momentum and frequency, we do not need to assume any kind of pairing mechanism a priori, 
but can numerically identify the channel which is dominant in the pairing interaction. This is 
a qualitative improvement over typical effective weak-coupling approaches where one channel (usually spin)
 is always assumed to dominate. As it will become clear later, we have obtained convincing numerical 
evidence that, at $t'/t\le0$, the spin component $\Phi_S$ not only dominates $\Gamma$ but also has
 the maximum value with the momentum transfer $\mathbf{Q}=(\pi,\pi)$, revealing that the superconducting 
pairing strength is mainly due to the antiferromagnetic fluctuations.

\begin{figure*}[ht!]
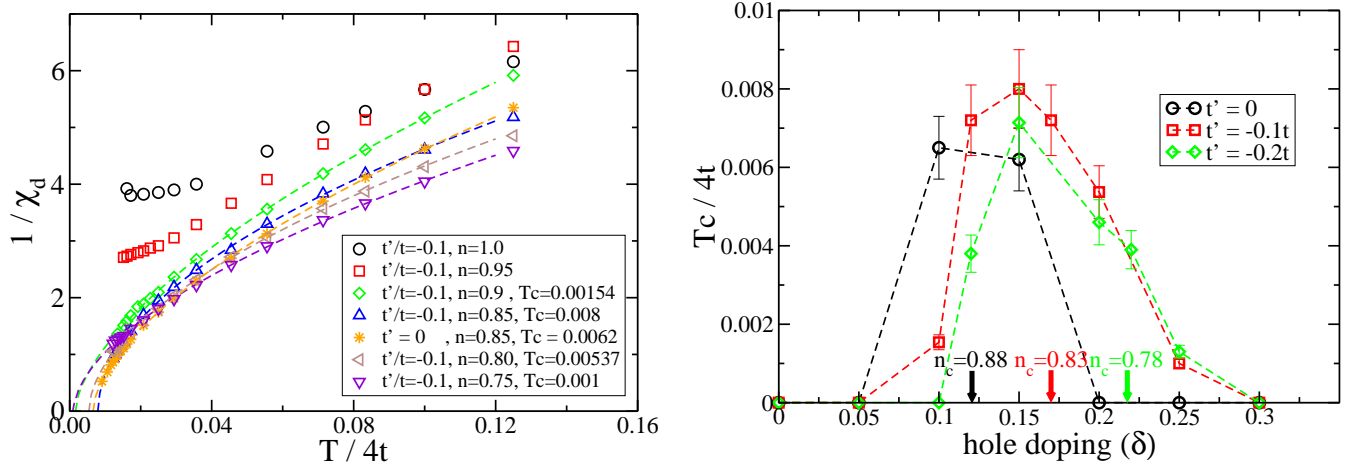

\centerline{
\includegraphics[width=3.4in]{ChixdT_12A16Bn_tpm0.1.eps} \quad
\includegraphics[width=3.4in]{Tc_combine_12A16B_ErrBar.eps}
}
\caption{ (color online) Left panel: $d$-wave pairing susceptibility
at $t'/t=-0.1$ and various doping concentrations (also one data set for $t'=0$ at $n=0.85$). 
The extrapolated transition temperature 
$T_c$ follows a dome shape curve between the fillings of $n=0.9$ (doping $\delta=0.1$) and $n=0.75$ (doping $\delta=0.25$), 
with its peak locates close to filling $n=0.85$. Right panel: Extrapolated $d$-wave superconducting 
transition temperature $T_c$ for $t'/t=0$, $-0.1$ and $-0.2$. The bold arrows indicate the corresponding quantum critical 
filling ($n_c$) along the Lifshitz line, determined in our previous work~\cite{Chen12}. The error bars of $T_c$ are obtained from 
the extrapolation of the $T_c$ for independent DCA/CTQMC simulations. The dome shifts towards 
higher doping as $t'/t$ becomes negative, but the peak of the dome is always located close 
to $\delta=0.15$. Results are obtained from $N_c=12$ and $16$ DCA/CTQMC simulations.}
\label{fig:Tc_tprime}
\end{figure*}

\section{Results}
\label{sec:results}

The left panel of Fig. \ref{fig:Tc_tprime} displays the $d$-wave pairing susceptibility for $t'/t=-0.1$
and doping concentration varying from half-filling to overdoped. Also, results for $t'=0$ at $n=0.85$ are shown. 
At or close to half-filling, $n=1.0$, $0.95$, where the system is inside the insulating or the pseudogap regions, the pairing 
susceptibility does not diverge at low temperatures. By increasing the doping beyond $n=0.9$ the pairing susceptibility 
becomes diverging at a finite $T_c$. The value of $T_c$ grows as the doping level changes from underdoped toward 
optimal doping at filling $n=0.85$. For higher doping, $T_c$ decreases and the $d$-wave superconducting dome eventually 
takes shape. Interestingly, above the optimal doping, we have observed marginal Fermi liquid behavior 
with self-energy linear on frequency and resistivity linear on temperature~\cite{Mikelsons2009,Chen11,Chen12}. 

\begin{figure}[h!]
\centering{}
\includegraphics[width=3.5in]{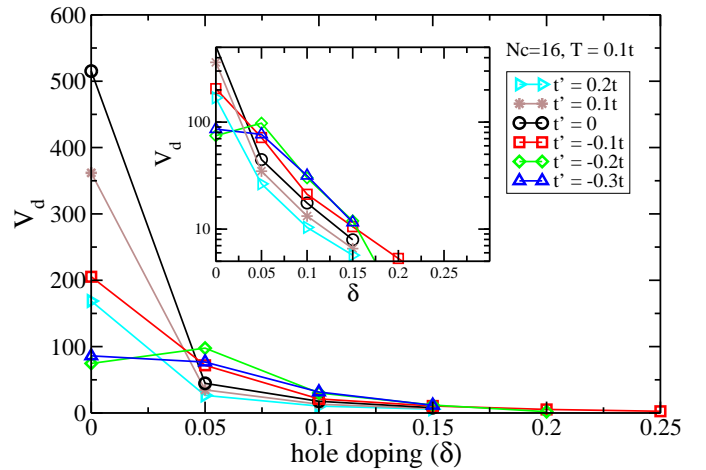}
\caption{ (color online) Evolution of the $d$-wave projected irreducible vertex function, $V_d$, 
as a function of hole-doping for various $t'/t$. At half-filling, $V_d$ is strongest at $t'=0$ 
and decays for both positive and negative $t'$. In the underdoped region (doping $\delta=0.05$, $0.1$ and $0.15$), 
$V_d$ is an increasing function as $t'$ varies from positive to negative. The inset presents the same data in a semi-log plot and the enhancement of $V_d$ from $t'>0$ to $t'<0$ can be clearly seen. 
 Also note that the curvature of the doping dependence of $V_d$ changes from convex to concave when $t'$ becomes 
negative. 
}
\label{fig:Vd_doping}
\end{figure}

The right panel of Fig.~\ref{fig:Tc_tprime} shows the extrapolated $T_c$ as function of doping 
for $t'/t=0$, $-0.1$ and $-0.2$. Two interesting observations can be made. First, as $t'/t$ becomes negative,
 the $d$-wave superconducting transition
 temperature increases, while the highest $T_c$ occurs around $n=0.85$. 
For more negative values of $t'/t$, e.g., $-0.3$, $-0.4$, $T_c$ decreases. Therefore, 
the superconducting dome evolves into a mountain-type volume in the $\delta$ vs $t'/t$ phase diagram. 
Second, we find that as $t'/t$ 
becomes more negative, the superconducting dome moves slightly towards higher doping. At first glance, this 
behavior seems to resemble the one of the Lifshitz line identified in Ref.~\onlinecite{Chen12}. 
The Lifshitz line traces those values of doping and $t'/t$ where the interaction-induced van Hove singularity 
crosses the Fermi level. Note we have pointed out the correspond quantum critical Lifshitz fillings, 
$n_c$, for $t'/t=0$, $-0.1$ and $-0.2$ by bold arrows in the right panel of Fig.~\ref{fig:Tc_tprime}. Naively, 
one would expect the Lifshitz filling to be associated with the maximum pairing interaction and transition temperature.
However, inspecting the behavior more quantitatively, the maximum of the dome actually stay close to n=0.85 
and does not fully follow  the Lifshitz line. 
We thus have to conclude that the interaction-induced van Hove singularity alone does not determine the 
superconducting transition, at least not the position of the highest transition temperature.

\begin{figure*}[ht!]
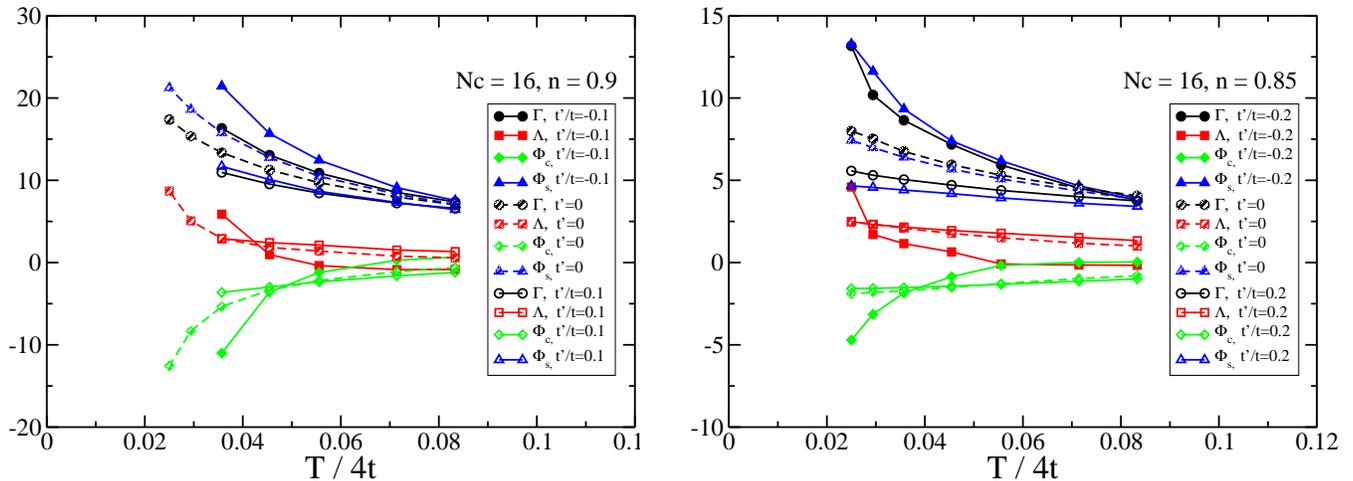

\centering{\includegraphics[width=3.4in]{vertexdecomposition_n0.9tp-0.2_n0.9tp-0.1_n0.9tp0_n0.9tp0.1_n0.9tp0.2.eps} \quad
\includegraphics[width=3.4in]{vertexdecomposition_n0.85tp-0.2_n0.85tp-0.1_n0.85tp0_n0.85tp0.2.eps}}
\caption{(color online) Decomposition of the irreducible pairing vertex $\Gamma$ into fully irreducible $\Lambda$, 
charge $\Phi_{C}$ and spin $\Phi_{S}$ contributions. (Left panel) Decomposition for filling $n=0.9$ 
(doping $\delta=0.1$) at $t'/t=-0.1,0$ and $0.1$. (Right panel) Decomposition for filling $n=0.85$ 
(doping $\delta=0.15$) at various $t'/t=-0.2, 0$ and $0.2$. In both cases, the $S=1$ magnetic components $\Phi_{S}$ are
the dominant contribution to the irreducible pairing vertex.
}
\label{fig:vertexdecomposition}
\end{figure*}

To understand this behavior, we explore the projected irreducible $d$-wave pairing vertex, $V_d$, as function of 
doping for various $t'/t$ using a $N_c=16$ site cluster and temperature $T=0.1t$ (Figure \ref{fig:Vd_doping}).  
Since $V_d$ is the effective pairing strength, the BCS condition $V_d \chi^{0}_d=1$ is roughly 
equivalent to the divergence of the $d$-wave pairing susceptibility~\cite{th_maier_06b,YangSX11}. In our previous 
work~\cite{YangSX11,Chen12}, we found that at $t'=0$ $V_d$ decays monotonically as a function of doping, 
forming a convex function. $\chi^{0}_{d}$ has an algebraic temperature dependence at the quantum critical filling 
and is much less sensitive to temperature at other fillings. 
The combined doping dependence of $V_d$ and $\chi^{0}_{d}$ gives rise to the superconducting dome at $t'=0$ with 
its peak located close to the quantum critical filling. Here, as shown in Fig.~\ref{fig:Vd_doping}, we find that 
$V_d$ develops a more complicated doping dependence for different values of $t'$. In general, it is largest at 
half-filling and decays to smaller values at larger doping. However, $V_d$ displays non-trivial features in the 
underdoped and optimally doped regions. The  inset to Fig. 3, where the data are presented in semi-log fashion, 
shows that for doping $\delta=0.05$, $0.1$, and $0.15$ the value of $V_d$ systematically increases when $t'/t$ 
goes from positive to negative values.  In particular, at $\delta=0.05$ and $t'/t=-0.2$, the value of $V_d$ is 
larger than that at half filling. 
The curvature of the doping dependence of the effective pairing strength, $V_d(\delta)$,  changes from a convex 
function for $t'\ge-0.1$ to a concave one for $t'<-0.1$.
However, the enhancement of $V_d$ 
stops 
beyond $t'=-0.3t$, and $V_d$ starts to strongly decrease  also as a function of $t'/t$.

Putting all these observations together, it is evident that $V_d$ develops a non-trivial landscape as function 
of $\delta$ and $t'/t$. At half-filling, $V_d$ is largest for $t'=0$, and decays as $|t'/t|$ increases. 
This behavior is easily understood as $t'$ introduces frustration 
into the system and suppresses the tendency towards antiferromagnetic order. This symmetry is destroyed for finite doping.
In the underdoped and optimally doped region, $V_d$ is enhanced 
as $t'/t$ goes from positive to negative values, and finally decays once $t'/t$ becomes very negative. Furthermore, the
doping dependence of $V_d$, changes from a convex to a concave function for negative $t'/t$. We think the
concave curvature of $V_d$ at negative $t'$ is responsible for the fixed position of the peak in the superconducting dome
at different $t'$. In addition, the faster decay of $V_d$ with doping at negative $t'$, due to its concave nature, actually
constraints the dome to be further pushed to higher doping, even though the Lifshitz line moves towards higher doping 
at negative $t'$. The different behaviors of $V_d$ for positive and negative $t'$ reflects the previous observation that in
lieu of superconductivity the states in those two regions are fundamentally different. For example for n=0.85, the system 
is a Fermi liquid state for $t'>0$, while it resides in the pseudogap phase for $t'<0$~\cite{Chen12}.

\begin{figure}[ht!]
\centering{}
\includegraphics[width=3.4in]{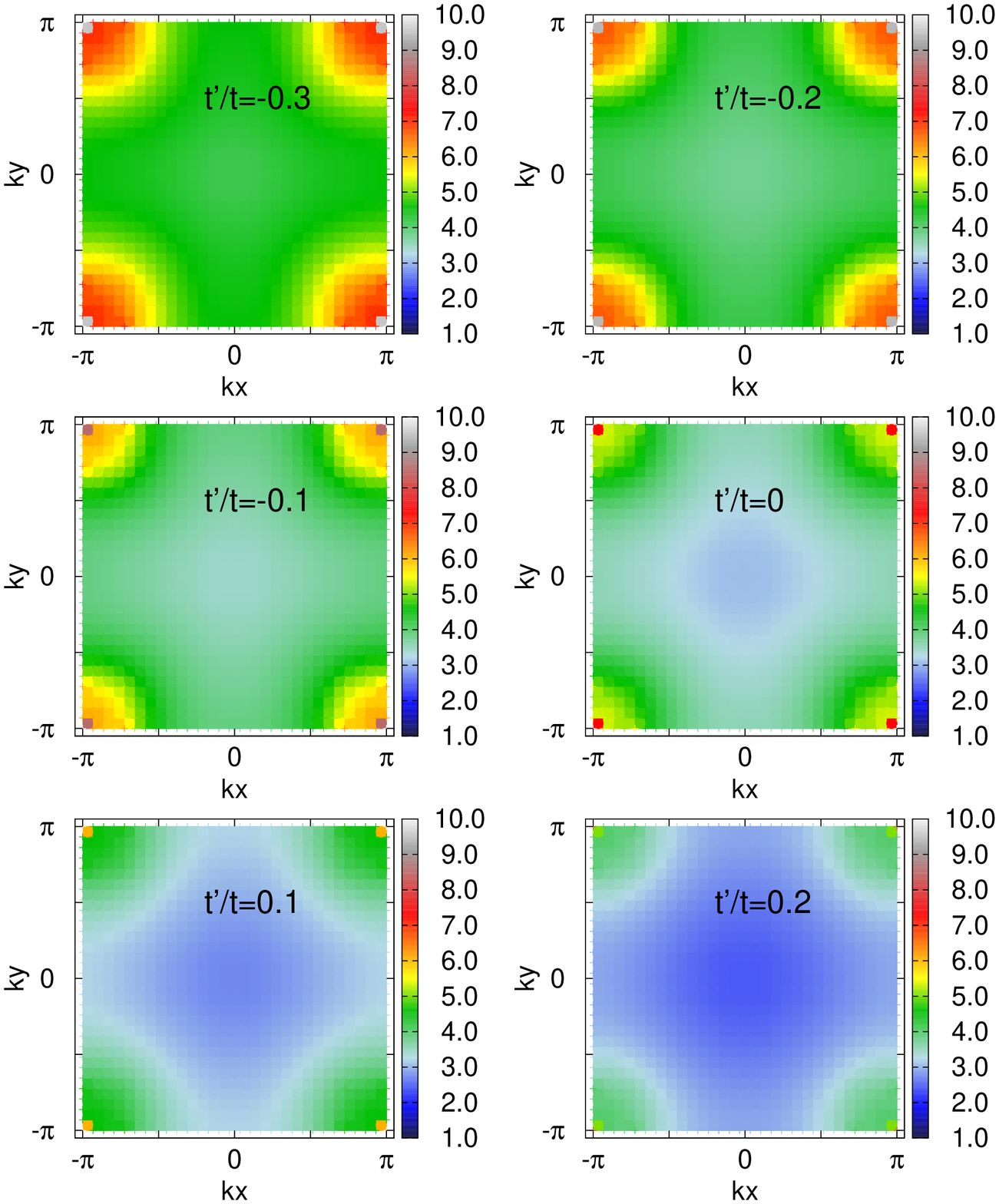}
\caption{(color online) Cluster spin susceptibility $\chi_{S}(\mathbf{Q},T)$ for a $N_c=16$ cluster, $n=0.85$ and $T=0.1t$. 
After interpolating the cluster spin susceptibility to the entire BZ, one clearly sees that as $t'/t$ goes from positive 
to negative values, the spin susceptibility develops greater weight at $\mathbf{Q}=(\pi,\pi)$,
corresponding to enhanced antiferromagnetic spin fluctuations at the negative $t'$ side of the phase diagram.
} 
\label{fig:ChiQT}
\end{figure}

To furthermore explore the pairing mechanism, we performed a decomposition of the irreducible 
pairing vertex $\Gamma$ via the parquet equations~\cite{Maier06,SXYang2009,Chen2013} as described in section II. The results are shown 
in Fig.~\ref{fig:vertexdecomposition}. The left panel is the vertex decomposition for filling $n=0.9$ 
and $t'/t=-0.1$, $0$ and $0.1$. 
The behavior of $\Gamma$ is consistent with the one of $V_d$ from Fig.~\ref{fig:Vd_doping}. 
For a fixed temperature, $\Gamma$ increases as $t'/t$ changes from $0.1$ to $-0.1$, and the 
increase becomes enhanced at lower temperatures. The dominant contribution to $\Gamma$ for $t'/t=-0.1$, $0$ and $0.1$
comes from the $S=1$ magnetic component $\Phi_S$. The same observations hold for the right panel of 
Fig.~\ref{fig:vertexdecomposition}, which corresponds to a filling $n=0.85$. In fact, we also find from the vertex decomposition that the momentum dependence of $\Phi_S$ is peaked at the momentum transfer $\mathbf{Q}=(\pi,\pi)$, associated with the antiferromagnetic fluctuations of the square lattice~\cite{Maier06}.

To clearly demonstrate the difference in the antiferromagnetic fluctuations at $t'>0$ and $t'<0$, 
we measure the cluster spin susceptibility $\chi_{S}(\mathbf{Q},T)$ for $N_c=16$, filling $n=0.85$ (doping $\delta=0.15$), 
and temperature $T=0.1t$, and we interpolate $\chi_{S}(\mathbf{Q},T)$ throughout the entire Brillouin Zone (BZ). The 
results are shown in Fig.~\ref{fig:ChiQT}. 
As one can see, $\chi_{S}(\mathbf{Q}=(\pi,\pi))$ becomes stronger as $t'/t$ goes from 
positive ($0.2$) to negative ($-0.3$) values, i.e. the antiferromagnetic fluctuations are indeed stronger for the negative 
$t'$ side. Together with the findings for the pairing strength $V_d$ and the decomposition of the irreducible pairing vertex 
$\Gamma$ into its different components, the behavior of the spin susceptibility makes the picture clearer. As mentioned 
previously, the system is a conventional Fermi liquid for $t'>0$, but in the pseudogap phase for $t'<0$. Obviously, 
the different physical properties are intimately connected with the spin fluctuations, 
as Fig. \ref{fig:ChiQT} shows. From Fig.~\ref{fig:vertexdecomposition} of our previous publication~\cite{Chen12}, we learned that the Fermi surface for $n=0.85$ is electron-like at 
$t'>0$ and hole-like at $t'<0$. The electron-like Fermi surface does not occupy the regions in the BZ close to $(\pi,0)$ and $(0,\pi)$, hence the system cannot
respond to the antiferromagnetic fluctuations; however, for the hole-like Fermi surface, regions close to $(\pi,0)$ and $(0,\pi)$ are occupied,
and the electrons there can couple to the antiferromagnetic fluctuation efficiently. From the vertex decomposition in Fig.~\ref{fig:vertexdecomposition} we also know that the antiferromagnetic fluctuations are strongest at the momentum transfer connecting $(\pi,0)$ and $(0,\pi)$. Hence, for $t'>0$ the system at optimal filling $n=0.85$ is a stable Fermi liquid, whereas for $t'<0$ it is already inside the pseudogap region
with electrons at the antinodal $(\pi,0)$ to $(0,\pi)$ momenta coupled strongly by antiferromagnetic fluctuations.

Previous studies~\cite{Honerkamp2001} have suggested that for strongly negative 
$t'$ ($>-0.5$) and close the so-called van-Hove filling, a ferromagnetic phase with a tendency 
towards p-wave superconductivity might appear. It surely is interesting to investigate whether this weak coupling prediction 
for large negative values of $t'$ is still valid in the strongly correlated system we study here.

\begin{figure}[t!]
\centering{}
\includegraphics[width=3.4in]{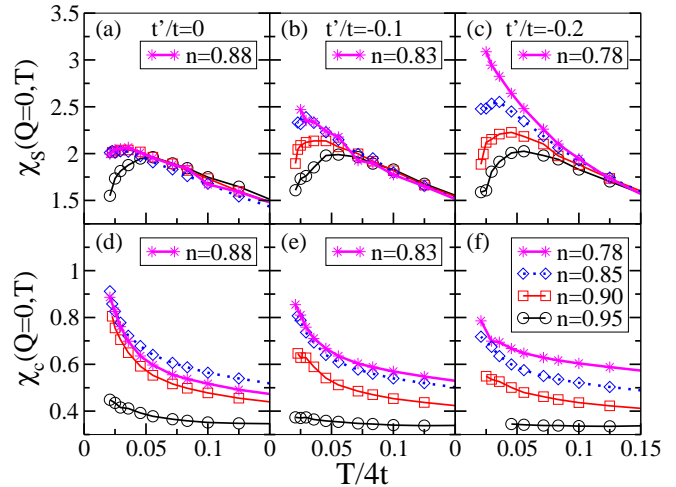}
\caption{(color online) (a), (b), (c) Bulk spin susceptibility ($\chi_{S}(\mathbf{Q}=0,T)$) vs temperature
 at $t'=0$, $t'/t=-0.1$ and $t'/t=-0.2$ and different fillings including  the corresponding quantum critical Lifshitz 
filling, $n_c(t'/t)$ (pink stars)
(d), (e), (f) Bulk charge susceptibility ($\chi_{C}(\mathbf{Q}=0,T)$) 
vs temperature for the same values of $t'/t$ and fillings. Results are obtained from $N_c=16$ DCA/CTQMC simulations. 
}
\label{fig:ChaChiT}
\end{figure}

To further probe our model we calculate the bulk spin and charge susceptibilities.
Fig.~\ref{fig:ChaChiT} (a), (b) and (c) show the bulk spin susceptibility, $\chi_{S}(\mathbf{Q}=0,T)$,
at different $t'/t=0$, $-0.1$ and $-0.2$ and fillings $n=0.95, 0.90, 0.85$ and the corresponding quantum critical Lifshitz 
filling at that value of $t'/t$, $n_c(t'/t)$. Panels (d), (e), and (f) display  $\chi_{S}(\mathbf{Q}=0,T)$
for the same values of $t'/t$ and filling. The maximum of the bulk spin susceptibility  is used to 
determine the pseudogap temperature $T^{\ast}$~\cite{Alloul1989,Vidhyadhiraja09,Sordi2012}. One can see in all 
the three cases, $T^{\ast}$ is highest close to 
half-filling ($n=0.95$), and gradually reduces as the doping increases. The lines with star symbols 
correspond to the Lifshitz filling~\cite{Chen12}, $n_c(t'/t)$, at which the Fermi surface changes its topology from 
electron-like to hole-like. As we have pointed out previously~\cite{Chen12}, the slope of $T^{\ast}$ 
as a function of doping $\delta$ becomes less steep as $t'$ goes to negative values, which means that the 
quantum critical region in the phase diagram becomes wider on the negative $t'$ side; it is also interesting
that for negative $t'$, $\chi_S(\mathbf{Q}=0,T)$ for the Lifshitz doping seems to diverge as
$\ln(1/T)$. If this  behavior is followed in an extended temperature region, it
would be consistent with a marginal Fermi liquid picture. The divergence of 
$\chi_S(\mathbf{Q}=0,T)$ in Fig.~\ref{fig:ChaChiT} (c), as well as the enhanced antiferromagnetic fluctuations 
observed in Fig.~\ref{fig:ChiQT}, convey the picture that as $t'$ goes from positive to negative values, 
the dominant fluctuations in the pseudogap and superconducting regions change in a very subtle manner. 
At $t'=0$, charge fluctuations are strongest close to the maximum of the superconducting dome, as evident from 
the diverging charge susceptibility, $\chi_C(\mathbf{Q}=0,T)$, in Fig.~\ref{fig:ChaChiT} (d). However, at $t'<0$, 
the charge susceptibility at all fillings including $n_c(t'/t)$ is suppressed, and at the same time the spin
susceptibility becomes more diverging. This observation is consistent with our previous results,~\cite{Khatami10,Chen12} 
where we showed that charge fluctuations indeed dominate in the vicinity of the finite temperature classical critical points for $t'>0$. 
However, as $t'$ becomes negative, charge fluctuations are suppressed in the doping region corresponding to the superconducting 
and pseudogap states. In these parameter regime, spin fluctuations become greatly enhanced, and from Fig.~\ref{fig:vertexdecomposition}
 we see the spin component dominates the pairing interaction. Such a change 
of dominant fluctuations from charge to spin, as $t'/t$ changes sign, is highly non-trivial 
and needs further analysis and understanding.

\section{Discussion}  
\label{sec:Discussion}

\begin{figure}[h!]
\centering{}
\includegraphics[width=3.4in]{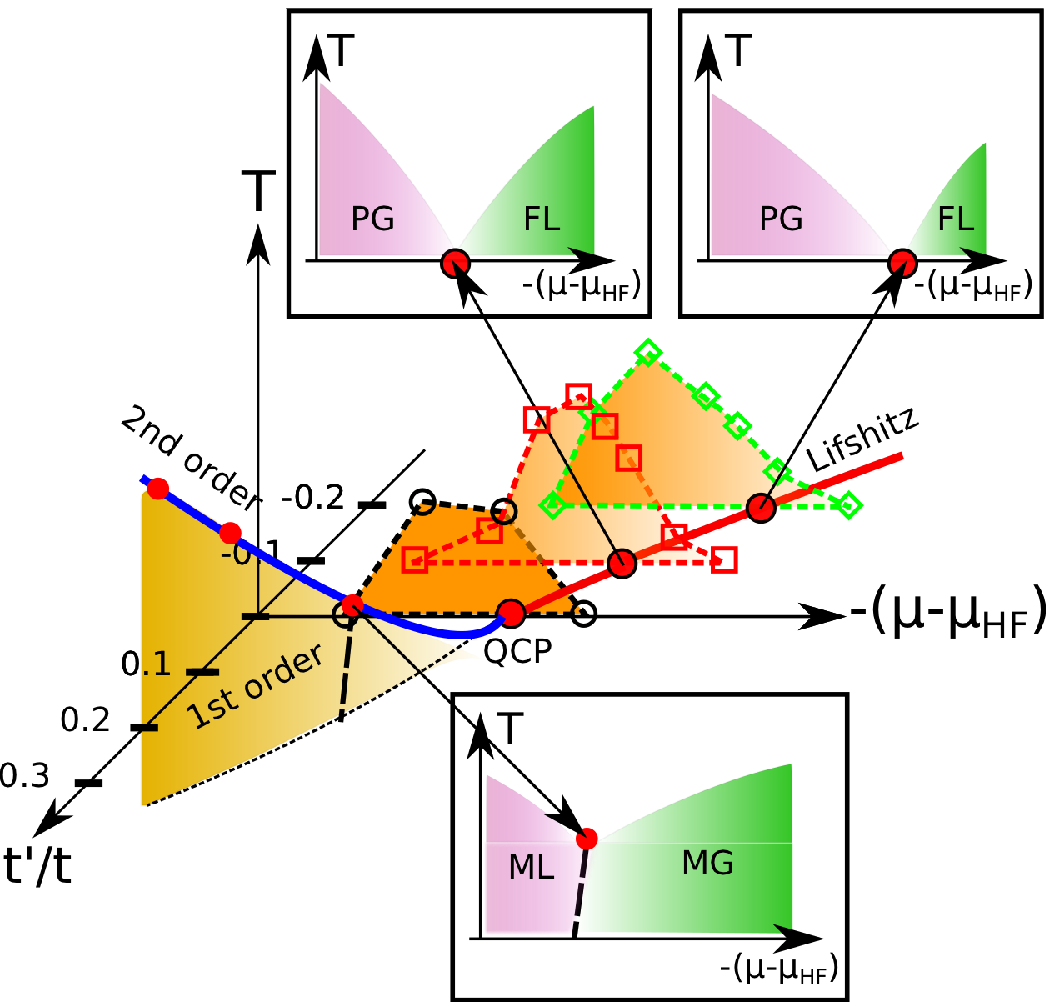}
\caption{(color online) Schematic phase diagram of the two-dimensional Hubbard model based on numerical data presented 
here and in previous publications, with temperature 
($T$), chemical potential ($\mu$), and next-nearest-neighbor hopping ($t'$) as the control parameters. 
For clarity we neglect antiferromagnetic phases. For each $t'$, we shift the chemical potential $\mu$ with 
respect to its half filling value $\mu_{\text{HF}}(t'/t)$. The three insets show the cut of 
the phase diagram at three different $t'/t=0.1,-0.1$ and $-0.2$. At positive $t'$, the Mott 
liquid (ML) and Mott gas (MG) phases are separated by a first order line; at negative $t'$, the quantum critical 
region separates the pseudogap (PG) and the Fermi liquid (FL) phases. This region  
gets wider as $t'$ becomes more negative. 
The superconducting dome, determined from the divergence of the $d$-wave pairing susceptibility, 
evolves with $t'/t$. At $t'=0$, the QCP roughly coincides with 
the center of the dome; however, as $t'$ goes to negative values, the shape of the dome becomes less symmetric, 
with its peak always locates close to filling $n\approx0.85$. However, the Lifshitz points move towards the overdoped 
boundary of the dome.}
\label{fig:PhaseDiagram}
\end{figure}

Quantum criticality is one of the driving forces behind many of the unconventional superconductors known today. 
Most prominent examples are the heavy fermions based on lanthanide or actinide inter-metallics~\cite{Stewart1984,vonloehneysen1997} 
and possibly also the iron pnictides~\cite{Walmsley2013,Shibauchi2013}. For the latter materials, the QCP has been unambiguously identified as driven by spin fluctuations, although the precise mechanism -- Hertz-Millis type spin-waves~\cite{Hertz1976,Millis1993} or Coleman-Si type local quantum criticality~\cite{Abrahams2011,Gegenwart2008} -- has not been unanimously settled yet.

The situation is even worse in the case of cuprate high-$T_c$ superconductors. Here, all
experimental evidence points towards a QCP just below the dome, but its nature is completely open. 
There are several competing scenarios. Varma, for example, favors a loop-current type QCP, 
based on the analysis of the marginal Fermi liquid in terms of symmetry arguments~\cite{He2012,Varma2013}. 
The success of the spin-fluctuation approach to explain both the superconducting dome and the anomalous features 
above it would point towards a spin-wave type QCP, but in the region of optimal doping there is no 
real evidence for something along that line. Others propose some hidden control parameter, 
driving the system off the QCP at the physically relevant values \cite{Sachdev2012a,Sachdev2012b,Vojta2009}. Finally, the Lifshitz 
transition scenario~\cite{Hackl2010}, possibly supported by the stripes sometimes observed \cite{Vojta2012}, is 
another candidate.

Based on our DCA/CTQMC simulations for the two-dimensional Hubbard model, we can support a combination of the latter two scenarios. 
As presented in the schematic phase diagram of Fig.~\ref{fig:PhaseDiagram}, 
for $t'>0$ we indeed find a line of phase transitions between a Mott liquid and a Mott gas. 
The interesting point is that the QCP for this phase  transition lies right at $t'=0$, and is continued by a line 
of Lifshitz points for $t'<0$, separating pseudogap and Fermi liquid regions. For not too large $t'<0$ the 
system  can still take advantage of the quantum fluctuations present in the vicinity of the QCP, which usually 
leads to superconductivity. The somewhat counter-intuitive observation however is, that the dome does not follow 
the Lifshitz line, but has a much weaker dependence in doping. Such a complicated behavior could however be understood 
by studying the pairing strength $V_d$ (see Fig.~\ref{fig:Vd_doping}). 
The curvature of the doping dependence of $V_d$ changes from convex to concave function as $t'$ changes its sign. 
This fixes the position of the peak of the superconducting dome to the same optimal filling of $n\approx0.85$
for all $t'$ studied. On the other hand, the faster decay of $V_d$ at higher doping due to its
concave curvature constraints the dome to extend
to higher doping at negative $t'$, although the Lifshitz line does so. 
Along the Lifshitz line, we furthermore observe that the spin fluctuations become 
enhanced over the charge fluctuations (see Fig.~\ref{fig:ChaChiT}), which are dominant at $t'\ge0$. 

We interpret this behavior in terms of the vicinity to the Mott liquid present at $t'>0$, which favors local moment 
formation and hence enhanced spin fluctuations in two-dimensions. For $t'>0$ superconductivity 
is inhibited by the presence of the Mott liquid, which is adiabatically connected to the pseudogap phase for $t'\le0$, 
where the superconductivity can survive longer (left of the Lifshitz line in  Fig. 7). 
This is due not only to the quantum critical fluctuations associated with the QCP, but a cooperation of incipient 
quasiparticle formation in the pseudogap phase together with surviving spin fluctuations from the latter. It is 
noteworthy that the Lifshitz line actually seems to be the line where superconductivity terminates or becomes 
dramatically suppressed; right to the Lifshitz line there are well-developed quasiparticle, but no strong enough 
fluctuations to  support superconductivity.

\section{Conclusion}
\label{sec:Conclusion}
Using large-scale dynamical cluster quantum Monte Carlo simulations, we map out the phase diagram of 
the two dimensional Hubbard model in the vicinity of the quantum critical filling. The control parameters of our simulation
are temperature $T$, chemical potential, and the next-nearest-neighbor hopping $t'/t$. We also include extensive 
two-particle measurements into the simulations to measure the $d$-wave pairing susceptibility directly. We can determine the 
superconducting transition temperature $T_c$ by extrapolating the inverse of the $d$-wave pairing susceptibility to zero. Then 
we find the superconducting dome is located around the quantum critical Lifshitz doping, and furthermore follow its evolution 
to the negative $t'$ side of the phase diagram. However, the evolution of the dome is subtle, 
since its peak stays close to the same filling ($n \approx 0.85$), whereas the whole dome moves  towards higher 
doping, following the trend of the Lifshitz line. The $d$-wave projected irreducible pairing vertex function or effective 
pairing strength is enhanced as $t'$ changes from positive to negative values in the under- and optimal-doped regions.
The doping dependence of the effective pairing strength, changes its curvature from convex to concave as $t'$ goes to negative 
values, which fixes the position of the maximum superconducting temperature at a filling of $n\approx0.85$ and further restricts 
the dome from following the Lifshitz line. The vertex decomposition via parquet 
equations furthermore reveals that in the negative $t'$ side of the phase diagram, the effective pairing strength is dominated by 
magnetic fluctuations associated with the antiferromagnetic order at momentum transfer $\mathbf{Q}=(\pi,\pi)$. Many interesting 
questions still remain, particularly, the reason of the changing curvature of the effective pairing strength $V_d$, and the 
change on the character of the dominant fluctuations from charge at $t'\ge0$ to spin at $t'<0$, 
and require further detailed investigations.

\begin{acknowledgments}
We would like to thank Ka-Ming Tam, Pablo Rodriguez, and Karlis Mikelsons for useful conversations. We thank Joseph Betouras, 
Jan Zannen, Henri Alloul, Zheng-Yu Weng and Alejandro Muramatsu for valuable discussions. 
This work is supported by NSF grants OISE-0952300 and DMR-0706379, and the NSF EPSCoR Cooperative Agreement No. 
EPS-1003897 with additional support from the Louisiana Board of Regents. TP acknowledges support by the 
DFG through the collaborative research center CRC 602 and the research unit FOR 1807. Supercomputer support was provided by the 
NSF Extreme Science and Engineering Discovery Environment (XSEDE) under grant number DMR100007, the Louisiana Optical Network 
Initiative, and HPC@LSU computing resources.
\end{acknowledgments}

\bibliography{Domevolution}

\end{document}